\documentclass[twocolumn,showpacs,aps,epsfig,show keywords,nofootinbib]{revtex4}

\usepackage{graphicx}
\usepackage{amsfonts}
\usepackage{epstopdf}
\usepackage{latexsym}
\usepackage{amssymb}
\usepackage{amssymb}
\usepackage{mathrsfs}

\usepackage[center]{subfigure}

\begin{document}

 \newcommand{\bq}{\begin{equation}}
 \newcommand{\eq}{\end{equation}}
 \newcommand{\bqn}{\begin{eqnarray}}
 \newcommand{\eqn}{\end{eqnarray}}
 \newcommand{\nb}{\nonumber}
 \newcommand{\lb}{\label}
\newcommand{\PRL}{Phys. Rev. Lett.}
\newcommand{\PL}{Phys. Lett.}
\newcommand{\PR}{Phys. Rev.}
\newcommand{\CQG}{Class. Quantum Grav.}

\title{Constructing analytical solutions of linear perturbations of inflation with modified dispersion relations}

\author{Tao Zhu$^{a,b}$, Anzhong Wang$^{a,b}$, Gerald Cleaver$^{c}$, Klaus Kirsten$^{d}$,  Qin Sheng$^{d}$}

 \affiliation{$^{a}$ Institute for Advanced Physics $\&$ Mathematics, Zhejiang University of Technology, Hangzhou, 310032,  China\\
$^{b}$  GCAP-CASPER, Physics Department, Baylor University, Waco, TX 76798-7316, USA \\
$^{c}$ EUCOS-CASPER, Physics Department, Baylor University, Waco, TX 76798-7316, USA\\
$^{d}$ GCAP-CASPER, Mathematics Department, Baylor University, Waco, TX 76798-7328, USA}

\date{\today}

\begin{abstract}

We develop a technique  to construct analytical  solutions of  the linear perturbations of inflation  with a nonlinear  dispersion relation, due to  quantum effects of the early universe. Error bounds are given and studied in detail. The analytical solutions describe the evolution of the perturbations extremely well even when only the first-order approximations are considered.   
\\
 

\end{abstract}


\maketitle
 
 \section{Introduction}
 
The inflationary cosmology \cite{Guth,Starobinsky,Sato} provides a framework for solving several fundamental and conceptual problems of the standard big bang cosmology \cite{cosmo}. Most importantly, it provides a causal mechanism for generating structures in the universe and the spectrum of cosmic microwave background (CMB) anisotropies.  These are matched to observations with unprecedented  precision \cite{WMAP,WMAP2}, especially after the recent release of the more precise results from the Planck satellite \cite{PLANCK}.

However, such successes  are contingent on the understanding of physics in much earlier epochs when temperatures and energies were much higher than what we are able to access elsewhere \cite{BCQ}. In particular,    if the inflationary period is sufficiently long,  the physical wavelength of fluctuations observed at the present time may well originate with a wavelength smaller than the Planck length at the beginning of the inflation  
- the trans-Planckian issues \cite{Brandenberger1999}. Then, questions arise as to whether the usual predictions of the scenario still remain robust, due to the 
ignorance of physics in such a small  scale, and more interestingly, whether they have left  imprints for future observations. 

Such considerations have attracted lots of   attention,  and various  approaches have been proposed\cite{Martin2001,Martin2002,Martin2003,Niemeyer2001,Ber,Martin2004,Jackson,Easther,eff,eff-2,eff-3,Brandenberger2013CQG}. One of them is to replace the  linear dispersion relation by a nonlinear one in the  equations of the perturbations. This approach was initially applied to inflation as a toy model \cite{Martin2001}, motivated from the studies of the dependence of black hole radiation on Planck scale physics \cite{UCJ,CJ}. Later, it was  naturally realized   \cite{HL,HL2,HL3,HL4,HL5,HL6,HL7,HL8,HL9,HL10} in the framework of the Ho\v{r}ava-Lifshitz gravity, a candidate of the ultraviolet complete theory of quantum gravity \cite{Horava,reviews,reviews2,reviews3}.

Hence, obtaining  approximate analytical solutions of the perturbations becomes one of the crucial steps  in understanding the quantum effects on inflation, including the power spectra of the perturbations, non-Gaussianities, primordial gravitational waves, temperature and polarization of CMB, and has been intensively investigated in the past decade \cite{Martin2001,Martin2002,Martin2003,Niemeyer2001,Ber,Martin2004,Jackson,eff,eff-2,eff-3,Brandenberger2013CQG,HL,HL2,HL3,HL4,HL5,HL6,HL7,HL8,HL9,HL10}. However, these studies  were carried out mainly   by using the  Brandenberger-Martin  (BM) method, in which the evolution of the  perturbations is divided into several epochs, and in each of them the approximate analytical  solution can be obtained either by  the WKB approximations when the adiabatic condition is satisfied, or by the linear combination of the exponentially decaying and growing modes,  otherwise. Then, the individual solutions were matched together at their  boundaries.  While this often yields reasonable analytical approximations, its validity  in more general cases  has been questioned recently, 
and shown that it is valid only when  $k \gg aH$  \cite{JM09}, where $k$ is the comoving wavenumber,  $a$  the expansion factor of the universe, and $H \equiv \dot{a}/a$ with $\dot{a} \equiv da/dt$. Therefore, unless the non-adiabatic evolution of the mode function is properly taken into account, the BM method cannot be applied to the case in which  the adiabatic condition of the evolution of the mode function is not guaranteed. In addition,  the errors are not known in this method.  However, with the arrival of the precision era of cosmological measurements, accurate calculations of cosmological variables are highly demanded \cite{KDM,KDM2}. 

In this paper, we propose another  method,  {\em the uniform asymptotic approximation}, to construct analytical solutions of the linear (scalar, vector or tensor) perturbations of inflation with modified dispersion relations. We construct   explicitly the  error  bounds and study them in detail. Because of the understanding and control of the errors, such constructed solutions describe the exact evolutions of the perturbations extremely well, 
even only in the first-order approximation [cf. Fig. \ref{fig1}]. 

It should be noted that the uniform asymptotic approximation  was first used to study the mode function  by  Habib {\em et al} \cite{uniformPRL,uniformPRD1}, and later applied to some
particular models \cite{KT08,KT}. 
However, their treatments are applicable only to the case where  the dispersion relation is linear $b_i =0$,  where $b_i$ are defined in Eq.(\ref{omega}), so that  $g(\eta) = 0$   
has only one   single root [cf. Eq.(\ref{functions})]. It cannot be applied to the more interesting cases with several roots, and in particular, 
to those where some roots may be double, triple or  even high-multipole roots. The method to be developed below shall treat all these cases in a unified way, 
which is mathematically quite different from that of \cite{uniformPRL,uniformPRD1}, and reduces to it when $b_i = 0$.

\section{Uniform Asymptotic Approximation}

In the slow-roll inflation, we have $a(\eta) \simeq - (1-\varepsilon)/(\eta H)$, with  $\eta$ and  
$\varepsilon \; [\equiv - \dot{H}/H^2]$ being, respectively, the conformal time and slow-roll parameter. Then,  the perturbations (of scalar, vector  or tensor)  
are given  by \cite{Brandenberger2013CQG},
\bqn
\lb{eom58}
\mu_{k}''(y)=\left[g(y)+q(y)\right]\mu_k(y),
\eqn 
where $y \equiv - k\eta$,  $\mu_k(y)$ denotes  the mode function,    a prime  the derivative with respect to $y$, and
\bq
\lb{eomb}
g(y)+q(y) \equiv \frac{\nu^2(y)-1/4}{y^{2}} - \hat\omega_k^2(y).
\eq
Here $\nu(y)$ depends on the background and types of perturbations. The modified dispersion relation $\hat\omega_k^2(y)$ takes the form, 
\bq
\lb{omega}
 \hat\omega^2_k(y) = 1-b_1\epsilon_*^2 y^2 + b_2 \epsilon_*^4 y^4,
 \eq
where $\epsilon_* \equiv H/M_*$, with 
$M_*$ being  the  energy scale, above which the quantum effects become important.   
To the first-order approximations of the slow-roll inflation, one can treat $\nu(y),\; H$ and  $b_i$ as
constants for all types of perturbations. For details, see for example \cite{Brandenberger2013CQG}.

Equation (\ref{eomb}) shows that $g(y)$ and $q(y)$ in general have two poles, at $y =  0^{+} $  and $y = +\infty$, respectively. In addition, $g(y)$ has multiple  turning points (or roots of the equation $g(y) = 0$). From the theory of the second-order linear differential equations, one finds that the asymptotic solutions of Eq.(\ref{eom58})  depend on the
behavior of $g(y)$ and $q(y)$ around the poles and turning points. Most of the approximate methods developed in the literature can yield good approximate solutions only in the regions that don't contain turning points \cite{Martin2001,Martin2002,Martin2003,Niemeyer2001,Ber,Martin2004,Jackson,uniformPRL,uniformPRD1}. The main purpose of this paper is to develop a uniform approach to construct analytical solutions of Eq.(\ref{eom58}) for $g(y)$ that can contain various poles and turning points. The essential and nontrivial feature of this approach is that it can address all these cases  in a unified way.

To proceed further,  let us first introduce the Liouville   
transformations with two new variables $U$ and $\xi$  \cite{Olver1974,Nayfeh},
\bqn
\lb{Olver trans}
&& U(\xi)= \chi^{1/4} \mu_k(y),\;\;
\chi=\frac{|g(y)|}{f^{(1)}(\xi)^2}=\left(\frac{d\xi}{dy}\right)^2,
\eqn
where 
\bq
\lb{ffunction}
f(\xi)=\int^y \sqrt{|g(\hat{y})|} d\hat{y},\;\;\; f^{(1)}(\xi)\equiv \frac{df(\xi)}{d\xi}.
\eq
In terms of $U$ and $\xi$,  Eq.(\ref{eom58}) takes the form,
\bqn
\lb{EoM}
\frac{d^2 U}{d\xi^2}=\Big[\pm f^{(1)}(\xi)^2+\psi(\xi)\Big]U,
\eqn
where 
\bq
\lb{psi}
\psi(\xi)\equiv \frac{q(y)}{\chi}-\chi^{-3/4} \frac{d^2(\chi^{-1/4})}{dy^2}, 
\eq
and the signs $\pm$ correspond to the cases of $g(y)>0$ and $g(y)<0$, respectively. Thus, to get the analytical solution of Eq.(\ref{eom58}), now we need to solve the new equation, i.e., Eq.(\ref{EoM}), which provides a better mathematical treatment near the turning points. In particular,  the following we shall show explicitly  how to determine the analytical approximate solutions of Eq.(\ref{EoM}) around the poles and turning points by properly choosing the function $f^{(1)}(\xi)^2$.

The essence of the approximation  is to choose properly a form of $f^{(1)}(\xi)^2$, which minimized the errors of the approximation, and meanwhile  enables  us to solve Eq.(\ref{EoM}) analytically in terms of known special functions. Obviously, the errors of the approximations is closely related to the amplitude of $\psi(\xi)$. As we will show below, this will be guaranteed by the requirements of the convergence of the error functions. On the other hand, from the Liouville transformation given in Eq.(\ref{Olver trans}),  one can see that $\chi$ is regular and doe not vanish in the intervals of interest. Consequently, $f(\xi)$ should be chosen so that $f^{(1)}(\xi)^2$ has zeros and singularities of the same types  as $g(y)$. 

Therefore,  the choices of $f^{(1)}(\xi)^2$ play an essential role in determining the accuracy of the approximate analytical solutions. 
Such a choice  crucially depends on the behavior of the function $g(y)$ near  the poles and turning points.
In particular,  in  the neighborhoods of the two poles,  the function $g(y)$ is regular, thus one has to choose $f^{(1)}(\xi)^2$ being regular in the same regions. The simplest choice is \footnote{In principle, one can choose any form of $f^{(1)}(\xi)^2$, as long as it is regular  in these regions. But one also needs to make sure that Eq.(\ref{EoM}) can be solved analytically with such a choice.}
$f^{(1)}(\xi)^2=\text{const}$.
Without loss of generality, we take this constant to be one. Then, we find that
\bq
\lb{xiA}
\xi=\int^y \sqrt{\pm g(\hat{y})} d\hat{y},
\eq
here $``\pm"$ correspond to the poles $y = 0^{+},\; +\infty$, respectively, and the equation of motion (\ref{EoM}) becomes
\bqn\lb{EoM-poles}
\frac{d^2 U}{d\xi^2}=\Big[\pm1+\psi(\xi)\Big]U.
\eqn
Then, to the first-order approximation, neglecting the $\psi(\xi)$ term in Eq.(\ref{EoM-poles}) we find
\bqn\lb{LG1}
\mu_k^{\pm}(y)&=& \frac{c_{\pm}}{[\pm g(y)]^{1/4}} e^{i^s \int^y \sqrt{\pm g(\hat{y})}d\hat{y}} \left(1+\epsilon^{\pm}_1\right) \nb\\
&& + \frac{d_{\pm}}{[\pm g(y)]^{1/4}} e^{-i^s \int^y \sqrt{\pm g(\hat{y})}d\hat{y}} \left(1+\epsilon^{\pm}_2\right),  ~~~~~
\eqn
where  $\epsilon^{\pm}_1$ and $\epsilon^{\pm}_2$ represent the errors of the asymptotic solutions, $c_{\pm}$ and $d_{\pm}$ are the integration constants,
and $s = 0$ ($s = 1$) at the pole $y = 0^+$ ($y = +\infty$).
The corresponding error bounds   
are given by \cite{Paper},
\bqn
&& |\epsilon^{+}_1|,\;\; \frac{1}{2} |g(y)|^{-1/2} \left|\frac{d\epsilon^{+}_1}{dy}\right| \leq e^{\frac{1}{2} \mathscr{V}_{0^+,y}(F)}-1,\nb\\
&& |\epsilon^{-}_1|,\;\; |g(y)|^{-1/2} \left|\frac{d\epsilon^{-}_1}{dy}\right| \leq e^{\mathscr{V}_{y,+\infty}(F)}-1,  
\eqn
where $\mathscr{V}_{x_1,x_2}(F) \equiv \int^{x_2}_{x_1} \left|{dF(y)}/{dy}\right|dy$,  and the
error control  function $\mathcal{F}(y)$  is defined as, 
\bq
\lb{Ffunction}
F(y) =  \int^y \left({|g|^{-1/4}}\frac{d^2}{dy^2}{|g|^{-1/4}}-{q}{|g|^{-1/2}}\right) d\hat{y}.
\eq
From the above expressions one can see that the errors sensitively depend on the choice of $g(y)$ and $q(y)$, and in order to minimize the errors one requires the error control function $F(y)$ to be convergent. As a crucial step of the approximate procedure, such requirement on the error control function is an essential condition to determine the splitting of functions $g(y)$ and $q(y)$.  To fix them uniquely, let us consider the above error bounds.
Let us first expand  
\bqn
\lb{expanA}
g(y)&=& y^{-m} \sum_{s=0}^\infty g_s y^{{s}},\;\;\;
q(y) =  y^{-n}\sum_{s=0}^\infty q_s y^{s},
\eqn
about $y = 0^+$.
Then,  the LG approximations are valid only when  $g(y)$ has a pole of order $m \geq 2$ \cite{Olver1974}. However, when $m > 2$, Eq.(\ref{eomb}) 
shows that  $|q(y)|<|g(y)|$ does not hold. Therefore, in the present case we must choose $m = 2$, for which the condition $|q(y)|<|g(y)|$ requires
$n \le 2$. Then, the convergence of $F(y)$ requires $n = 2,\; q_0 = -1/4$, while the condition $|q(y)|<|g(y)|$ leads to $q_1 = g_1 = 0$. On the other hand, at the pole
$y = \infty$, we can make similar expansions, i.e.,  
\bqn
\lb{expan}
g(y)&=& y^{\bar{m}}\sum_{s=0}^\infty \bar{g}_s y^{{-s}},\;\;\;
q(y) =  y^{\bar{n}}\sum_{s=0}^\infty \bar{q}_s y^{-s}.
\eqn
Following similar arguments given at $y = 0^+$, we find that $\bar{m} = 4,\; \bar{n} < 1$. Thus, $q(y)$ must take the form, $q(y) = -1/(4y^2) + q_2$, 
where $q_2  = -(1+g_2)$ and $|q_2| < |g_2|$. Then, without loss of generality, we can always set $q_2 = 0$ and finally obtain  \cite{Paper},
\bqn
\lb{functions}
q(y) &=& -\frac{1}{4y^2},\nb\\
 g(y)&=& \frac{\nu^2}{y^2}-1+b_1 \epsilon_*^2 y^2 -b_2 \epsilon_*^4 y^4.
\eqn

The LG approximate solutions presented in the above are valid only in the region where $g(y) \not= 0$. Once $g(y)$ are zero,
both $\psi(\xi)$ and $\mu_k(y)$ diverge, and the LG approximations become invalid. In order to get the asymptotic solutions
around these turning points, we need to choose  a different  $f^{(1)}(\xi)^2$ in Eq.(\ref{Olver trans}). However,
such choice depends on the nature of the turning points. But nevertheless, since $g(y)=0$ is in general a cubic equation,
it can be always cast in  the form
\bq
\lb{2.a}
b_2 x^6 -b_1 x^4+x^2-\nu^2 \epsilon_*^2=0,
\eq
where $x=\epsilon_* y$.
Let  
\bq
\lb{2.b}
\Delta\equiv ({\cal{Y}}-1)^3+\frac{1}{4}\left(2- {3} {\cal{Y}}+ 3 b_1 {\cal{Y}}^2 \nu^2 \epsilon_*^2\right)^2,
\eq
and   ${\cal{Y}}\equiv 3 b_2/ b_1^2$.
Then,  when $\Delta < 0$, there exist three distinct real single roots, denoted by $y_{i}\;(i = 0, 1, 2)$, respectively.
Without loss of generality, we further assume  $y_0<y_1 < y_2$. When $\Delta = 0$, we have one real single root $y_0$, and one real
double root $y_1 = y_2$, with $y_0 < y_1$. However, in this case it is impossible to have all  three roots equal. 
When $\Delta > 0$, there exists  only one real single root $y_0$, while $y_1$ and $y_2$ become single complex roots with $y_1 = y_2^{*}$. In
all the three cases, we have $y_0 \sim {\cal{O}}(1)$, while
the magnitudes of the roots $y_2$ and $y_1$ depend on $\epsilon_*$. Physically, we expect $\epsilon_*\ll1$. Then, we
have $y_{1},~y_{2}\gg 1$ for the cases $\Delta \le 0$. However,   the difference between $y_{1}$ and $y_{2}$ generally depends on the choice of 
$b_i$ and $\epsilon_*$.  

To process further, let us  consider  the three conditions  \cite{Olver1974,Nayfeh,Olver1975,Zhang1991}:  
{(a)} $|q(y)|< |g(y)|$ in regions except for the neighborhoods of $y_{i} (i = 0, 1, 2)$;
{(b)} $|q(y)|<|g(y)/(y-y_i)|$ in the neighborhoods of $y_{i}$;   and {(c)} $|q(y)|<|g(y)/[(y-y_1)(y-y_2)]|$ in the neighborhoods of $y_1\;\mbox{and}\;y_2$.

In the neighborhood of $y_0$, conditions (a) and (b) are satisfied, and   $y_0$ can be treated as a simple turning point. Then,  we can
 introduce a monotone increasing or decreasing function $\xi$ as
$f^{(1)}(\xi)^2=\pm\xi$,
where $\xi (y_0)=0$. Without loss of generality, we can always choose $\xi$ to have  the same sign as $g(y)$, and thus $\xi$ is a
monotone decreasing function around  $y_0$ and is given by
\bqn
\xi=\cases{
                   - \left(\frac{3}{2} \int^y_{y_0} \sqrt{-g(\hat{y})} d\hat{y}\right)^{\frac{2}{3}},& $ y\geq y_0$,\cr
                   \left(-\frac{3}{2} \int^y_{y_0} \sqrt{g(\hat{y})} d\hat{y}\right)^{\frac{2}{3}},& $ y\leq y_0$.\cr}
\eqn
The equation of motion (\ref{EoM}) now becomes
\bqn\lb{EoMxi}
\frac{d^2U(\xi)}{d\xi^2}=[\xi+\psi(\xi)]U(\xi).
\eqn
Neglecting the $\psi(\xi)$ term in the first-order approximation of Eq.(\ref{EoMxi}), we obtain 
\bqn
U(\xi)=\alpha_0 \Big(\text{Ai}(\xi) +\epsilon_3(y)\Big) +\beta_0 \Big( \text{Bi}(\xi) +\epsilon_4(y)\Big),
\eqn
where $\text{Ai}(\xi)$ and $\text{Bi}(\xi)$ are the Airy functions, and $\epsilon_{3, 4}(y)$    denote the errors of the approximate solutions. Then the mode function is expressed as
\bqn\lb{airy solution}
\mu_k(y)&=& \alpha_0  \left(\frac{\xi}{g(y)}\right)^{1/4}\Big(\text{Ai}(\xi) +\epsilon_3(y)\Big)\nb\\
&&+\beta_0 \left(\frac{\xi}{g(y)}\right)^{1/4}\Big( \text{Bi}(\xi) +\epsilon_4(y)\Big),
\eqn
In particular,  
the error bounds now   
read  \cite{Paper},
\bqn\lb{error0}
\frac{|\epsilon_3|}{M(\xi)},\;\frac{|\partial \epsilon_3/\partial \xi|}{N(\xi)} \leq \frac{E^{-1}(\xi)}{\lambda} \Bigg[\exp{\Bigg\{\lambda \mathscr{V}_{\xi,a_3}(\mathscr{H})\Bigg\}}-1\Bigg],\nb\\
\frac{|\epsilon_4|}{M(\xi)},\;\frac{|\partial \epsilon_4/\partial \xi|}{N(\xi)} \leq \frac{E(\xi)}{\lambda} \Bigg[\exp{\Bigg\{\lambda \mathscr{V}_{a_4,\xi}(\mathscr{H})\Bigg\}}-1\Bigg],\nb\\
\eqn
where $\xi\in[a_4,a_3]$, $\mathscr{H}(\xi)$ denotes the corresponding error control function, given by 
\bq
\lb{Hfunction}
\mathscr{H}(\xi)=\int^{\xi} |v|^{-1/2} \psi(v) dv,
\eq
 and $M(\xi), \; N(\xi)$, and $\lambda$ are given explicitly in \cite{Olver1974a,Paper}.

Near the turning points $y_1$ and $y_2$, conditions (a) and (c) are always satisfied.  
When condition (b) is also satisfied,  we can treat $y_1$ and $y_2$ as single  turning points, and similar to $y_0$, 
we can get the asymptotic solutions near them. However, when $y_2 - y_1 \ll 1$, condition (b) is not satisfied, and  the method used for   $y_0$ is no longer valid. 
Following Olver \cite{Olver1975}, we adopt  a  method to treat all these cases together.  The crucial step is to the choice
 $f^{(1)}(\xi)^2=|\xi^2-\xi_0^2|$, where $\xi$ is an increasing variable and $\xi(y_2) = - \xi(y_1)=\xi_0$. The case $y_1 = y_2$ corresponds to $\xi_0=0$,
and the one with a pair of complex conjugate roots   corresponds to $\xi_0^2 < 0$. Then, $\xi(y)$ is given by  
\bqn
\lb{xib}
\int^{y} \sqrt{|g(\hat{y})|}\; d\hat{y}=\int^\xi \sqrt{|{\xi'}^2-\xi_0^2|}\; d\xi',
\eqn
where
$\xi_0^2=\pm |({2}/{\pi}) \int_{y_1}^{y_2} \sqrt{g(y)} dy|$,  
and ``+'' (``-'')  corresponds to real (complex) $y_{1,2}$.
Thus, we obtain
\bqn\lb{solutionW}
&& \mu_k(y) = \alpha_1 \left(\frac{\xi^2-\xi_0^2}{-g(y)}\right)^{1/4} \left[W\left(\frac{1}{2}\xi_0^2, \sqrt{2}\xi \right) + \epsilon_{5}(\xi)\right]\nb\\
&&+\beta_1 \left(\frac{\xi^2-\xi_0^2}{-g(y)}\right)^{1/4} \left[W\left(\frac{1}{2}\xi_0^2, -\sqrt{2}\xi \right)+ \epsilon_{6}(\xi)\right],~
\eqn
where $W\left(\frac{1}{2}\xi_0^2, \pm \sqrt{2}\xi\right)$ are the parabolic cylinder functions, with $\xi$ now being given by Eq.(\ref{xib}).
Then,  the error bounds read \cite{Paper},
\bqn
&&\frac{|\epsilon_5|}{M\left(\frac{1}{2}\xi_0^2,\sqrt{2} \xi\right)},\;\frac{|\partial \epsilon_5/\partial \xi|}{\sqrt{2} N\left(\frac{1}{2}\xi_0^2,\sqrt{2}\xi\right)}\nb\\
&&\;\;\;\;\leq \frac{\kappa}{\lambda E\left(\frac{1}{2}\xi_0^2,\sqrt{2}\xi\right)} \Bigg[\exp{\Bigg\{\lambda \mathscr{V}_{\xi,a_5}(I)\Bigg\}}-1\Bigg],\nb\\
&&\frac{|\epsilon_6|}{M\left(\frac{1}{2}\xi_0^2,\sqrt{2} \xi\right)},\;\frac{|\partial \epsilon_6/\partial \xi|}{\sqrt{2} N\left(\frac{1}{2}\xi_0^2,\sqrt{2} \xi\right)}\nb\\
&&\;\;\;\;\leq \frac{\kappa E\left(\frac{1}{2}\xi_0^2,\sqrt{2} \xi\right)}{\lambda} \Bigg[\exp{\Bigg\{\lambda \mathscr{V}_{0,\xi}(I)\Bigg\}}-1\Bigg],\nb\\
\eqn
for $\xi>0$, where $a_5$ is the upper bound of $\xi$, $I(\xi)$ is the error control function, now defined as 
\bq
\lb{Ifunbction}
I(\xi)=\int^\xi |v|^{-1} \psi(v) dv,
\eq
and $\lambda$, $M\left(\frac{1}{2}\xi_0^2,\sqrt{2} \xi\right)$, $\kappa$, $N\left(\frac{1}{2}\xi_0^2,\sqrt{2} \xi\right)$, 
and $E\left(\frac{1}{2}\xi_0^2,\sqrt{2} \xi\right)$ are given explicitly in \cite{Olver1975,Paper}. It is easy to get the error bounds for $\xi<0$ by replacing the above $\xi$ by $-\xi$.

So far, using the Liouville transformations, we have found the analytical approximate solutions near the poles $y = 0^+, \; \infty$, given by Eq.\  (\ref{LG1}),  and in the neighborhoods of the
turning points $y_{i}$, given, respectively, by Eqs.~(\ref{airy solution}) and (\ref{solutionW}). We now move onto  determining the integration constants from the initial conditions
by matching them on their boundaries.
In this paper, we assume that the universe was initially at the adiabatic vacuum \cite{cosmo,Brandenberger2013CQG},
\bqn
\lim_{y\rightarrow+\infty}{\mu_k(y)} &=& \frac{1}{\sqrt{2 \omega_k(\eta)}} e^{-i \int^{\eta}\omega_k(\hat{\eta}) d\hat{\eta}},
\eqn
where $\mu_k(y)$ also satisfies   the Wronskian,
\bq
\lb{W}
\mu_k(y) \mu^{*}_k(y)'-\mu^{*}_k(y)\mu_k'(y)=i.
\eq
Applying them   to the solution $\mu^{-}_{k}(y)$, we find
\bqn
\lb{IC}
c_{-}=0,\;\;\;\;d_{-}=\frac{1}{\sqrt{2k}}.  
\eqn
 
On the other hand, to consider  the matching between the solution $\mu^{-}_{k}(y)$  and the one given by Eq.\  (\ref{solutionW}), we note that
for  positive and large $\xi$ the cylindrical functions take the asymptotic forms \cite{Gil2004},
\bqn\lb{asyW}
&&W\left(\frac{1}{2}\xi_0^2,\sqrt{2}\xi\right)\simeq \left(\frac{2 j^2(\xi_0)}{\xi^2-\xi_0^2}\right)^{1/4}
\cos{\mathfrak{D}},\nb\\
&&W\left(\frac{1}{2}\xi_0^2,-\sqrt{2}\xi\right)\simeq \left(\frac{2 j^{-2}(\xi_0) }{\xi^2-\xi_0^2 }\right)^{1/4}  \sin{\mathfrak{D}}, 
\eqn
where $j(\xi_0)\equiv \sqrt{1+e^{\pi \xi_0^2}}-\sqrt{e^{\pi \xi_0^2}}$, and
\bqn
\mathfrak{D}& \equiv& \frac{1}{2}\xi \sqrt{\xi^2-\xi_0^2}- \frac{1}{2}\xi_0^2 \ln{\left(\xi+\sqrt{\xi^2-\xi_0^2}\right)}\nb\\
&&+\frac{1}{2}\xi^2_0 \ln{|\xi_0|}+\frac{\pi}{4}+\phi\left(\frac{1}{2}\xi_0^2\right), 
\eqn
with 
\bq
\lb{phi}
\phi(x)\equiv\frac{x}{2}-\frac{x}{4} \ln{x^2}+\frac{1}{2}\text{ph}\Gamma\left(\frac{1}{2}+i x\right),
\eq
where  the phase $\text{ph}\Gamma\left(\frac{1}{2}+i x\right)$ is zero when $x=0$, and  determined  by 
continuity, otherwise. Inserting the above into Eq.\ (\ref{solutionW}), and then comparing the resulting solution with $\mu^{-}_{k}(y)$,
we find that continuity of the mode function and its first derivative with respect to $\eta$ leads to,
\bqn
\lb{ab1}
\alpha_1=   \frac{k^{-1/2}}{2^{3/4}} j(\xi_0)^{-1/2}, \;\;\;  
\beta_1 = -i   \frac{k^{-1/2}}{2^{3/4}} j(\xi_0)^{1/2}. 
\eqn

\begin{figure}[t]
\centering
	{\includegraphics[width=75mm]{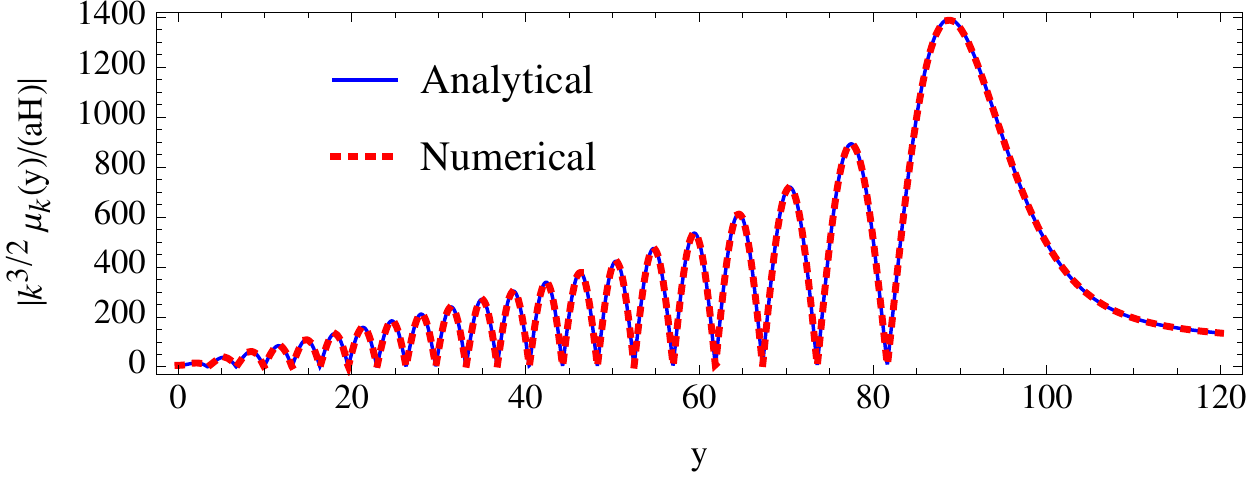}}
	{\includegraphics[width=75mm]{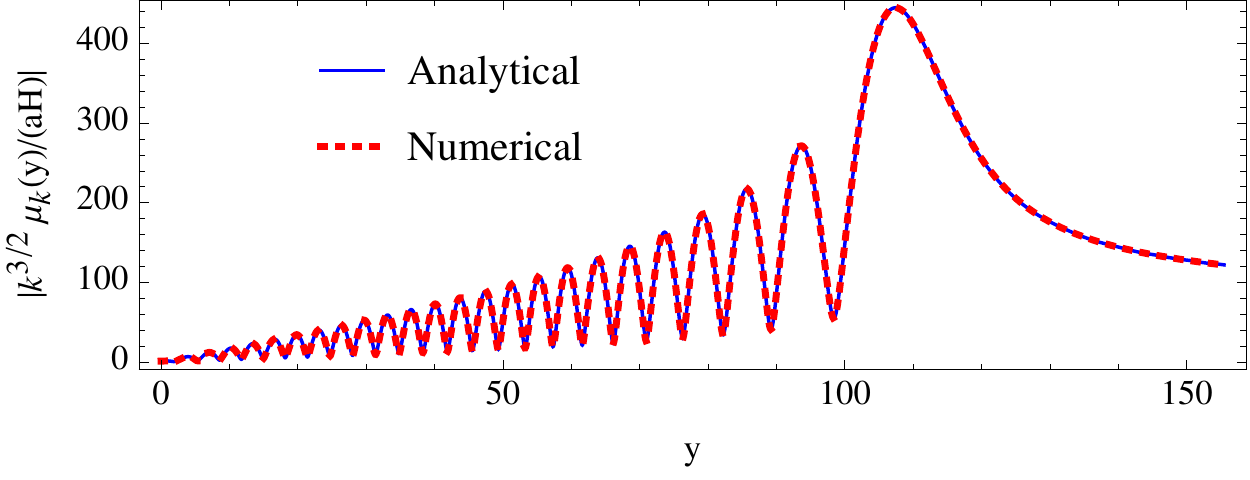}}
	{\includegraphics[width=75mm]{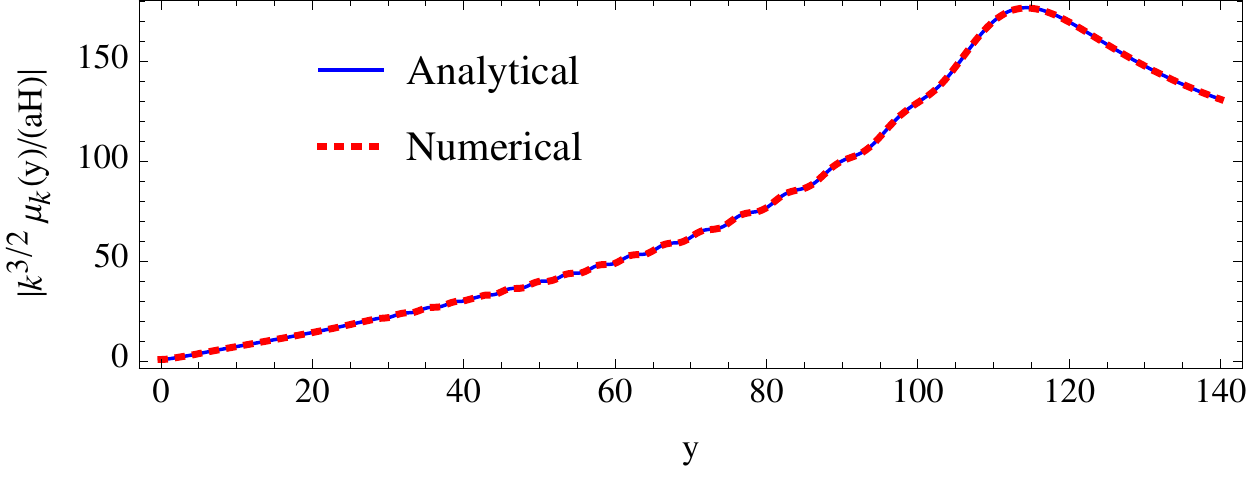}}
\caption{The numerical (exact)  (red dotted curves) and analytical (blue solid curves) solutions: 
(a) Top panel: Three  turning points  with  $b_1=2$, $b_2=0.98$. 
(b) Midlle panel: Two  turning points  with $b_1=1.5$, $b_2=0.5625949579339$. 
(c) Low panel: One turning point  with $b_1=1.5$, $b_2=0.59$.
In all three cases, we have set $\nu=3/2$ and $\epsilon_*=0.01$.}
\lb{fig1}
\end{figure}

To determine the coefficients $\alpha_0$ and $\beta_0$, we match the solutions (\ref{airy solution})
and (\ref{solutionW}) within the region
 $y\in (y_0,y_1)$. In this region,  $|y_0-y_1|$ is  very large, as mentioned above, and  $\xi$ is  
negative. Thus, from the asymptotic formula (\ref{asyW}) of $W\left(\frac{1}{2}\xi_0^2,\sqrt{2}\xi\right)$, and
the asymptotic form of the Airy functions,
\bqn\lb{negative large}
\text{Ai}(-x)&=&\frac{1}{\pi^{1/2} x^{1/4}}\cos{\left(\frac{2}{3}x^{3/2}-\frac{\pi}{4}\right)},\nb\\
\text{Bi}(-x)&=&-\frac{1}{\pi^{1/2} x^{1/4}}\sin{\left(\frac{2}{3}x^{3/2}-\frac{\pi}{4}\right)},
\eqn
 for $x \gg 1$, we find 
\bqn\lb{coevv}
\alpha_0&=& \sqrt{\frac{\pi}{2k}}\; \left[j^{-1}(\xi_0) \sin{\mathfrak{B}} - i \cdot  j(\xi_0) \cos{\mathfrak{B}}\right],\nb\\
\beta_0&=& \sqrt{\frac{\pi}{2k}}\; \left[j^{-1}(\xi_0) \cos{\mathfrak{B}} + i \cdot  j(\xi_0) \sin{\mathfrak{B}}\right],
\eqn
where 
\bq
\lb{2.d}
\mathfrak{B} \equiv \int_{y_0}^{y_1} \sqrt{-g}dy+\phi(\xi_0^2/2).
\eq
Finally, we consider the matching between $\mu^{+}_{k}(y)$ given by Eq.\   (\ref{LG1}) and the one given by
Eq.\   (\ref{airy solution}) in the region $y \in (0, y_0)$.
It can be shown that the continuity condition  yields
\bqn
\lb{cdp}
d_{+} &=& \left(\frac{\alpha_0\beta_0}{2\pi }\right) c_{+}^{-1} = \frac{\alpha_0}{2 \sqrt{\pi}} \exp{\left(-\int_{0^+}^{y_0} \sqrt{g}dy\right)}. ~~~
\eqn

Once we have uniquely determined the integration constants from the initial conditions,  
let us turn to consider some representative cases. In particular,  in   the case with three different single turning points, the numerical
(exact) and our  analytical approximate solutions are plotted in Fig. \ref{fig1}(a). The cases with two and one turning point(s) are plotted, 
respectively, in  Figs. \ref{fig1}(b) and \ref{fig1}(c).
From these figures, one can see  how well  the exact solutions are approximated by  our analytical ones. In fact, we have considered
many other cases, and found that in all those cases the exact solutions are extremely well approximated by the analytical ones.

\section{ Conclusions}

 In this paper, we have proposed a new method to construct analytical solutions  of linear perturbations of inflation,  
in which the dispersion relations are generically nonlinear and include high-order momentum terms, due to the quantum effects of the  early universe. 
The explicit  error  bounds are constructed for the error terms associated with the approximations. As a consequence, the errors   are well understood 
and controlled, and the analytical  solutions describe the exact evolution of the perturbations extremely well even in the  first-order approximation, 
as shown   in Fig. \ref{fig1}.   Thus, with this method it is expected that the accuracy of the calculations of cosmological variables, such as the power spectra, 
non-Gaussianity,  primordial gravitational waves,  temperature and polarization of  CMB, shall be significantly improved.

It should be noted that,  although in this paper we have  considered only the case where the maximal number of roots of the equation 
$g(y) = 0$  is three, our method can be easily generalized to the 
case with any number, and each of which can be a single, double, triple or even  higher multiple root. An interesting case is 
when the parity is violated in the early universe \cite{cosmoLV,cosmo-2,cosmo-3,cosmo-4,cosmo-5}, and terms like the Chern-Simons and fifth-order derivatives appear \cite{HL,HL2,HL3,HL4,HL5,HL6,HL7,HL8,HL9,HL10}. 

In addition, high-order approximations can also be constructed \cite{Olver1974,uniformPRL,uniformPRD1,Nayfeh}.  All these can be done 
not only in the case within the slow-roll inflation considered above, but also in more general inflationary backgrounds. 

Moreover, the case $b_i = 0$ was studied extensively
by using various methods, including the Green-function one \cite{SG01}. It would be very interesting to compare the results obtained by our method and the ones 
obtained by others. Such considerations are clearly out of the scope of this paper, and we hope to address them in another occasion.

\section*{Acknowledgements}
We thank Yongqing Huang, Jiro Soda,  Qiang Wu, and Wen Zhao for valuable discussions/comments. This work is supported in part by DOE, DE-FG02-10ER41692 (AW), Ci\^encia Sem Fronteiras, No. 004/2013 - DRI/CAPES (AW), NSFC Nos: 11375153 (AW), 11173021(AW), 11047008 (TZ), 11105120 (TZ), and  11205133 (TZ).

\end{document}